\documentclass[aip,graphicx, twocolumn, reprint]{revtex4-2}

\usepackage{amsmath}
\usepackage{amssymb}
\usepackage{styles}
\usepackage{flushend}
\usepackage{tabularx}

\setcitestyle{maxnames=3}

\usepackage{siunitx}

\newcommand{\dims}[4]{(\numproduct{#1 x #2 x #3})\,\unit{#4}}

\newcommand{\nozzleLength}{\qty{3}{\milli\meter}}
\newcommand{\nozzleDiameter}{\qty{0.3}{\milli\meter}}

\newcommand{\ovenDistance}{\qty{0.1}{\meter}}

\newcommand{\ovenAngleLimitDegree}{\qty{14}{\degree}}

\newcommand{\ovenProbePower}{\qty{9}{\micro\watt}}
\newcommand{\ovenProbeDiameter}{\qty{480}{\micro\meter}}

\newcommand{\zsPower}{\qty{87}{\milli\watt}}
\newcommand{\zsDetuning}{\qty{300}{\mega\hertz}}
\newcommand{\zsBeamWaist}{\qty{3.2}{\milli\meter}}

\newcommand{\threeDMOTRadius}{\qty{10}{\milli\meter}}
\newcommand{\threeDMOTPower}{\qty{3}{\milli\watt}}
\newcommand{\threeDMOTDetuning}{\qty{-18}{\mega\hertz}}

\newcommand{\threeDMOTMagnetDims}{\dims{38}{11}{7}{\milli\meter}} 

\newcommand{\icl}{\affiliation{Department of Physics, Blackett Laboratory, Imperial College London,
Prince Consort Road, London, SW7 2AZ, UK.}}
\newcommand{\ral}{\affiliation{Rutherford Appleton Laboratory, UKRI--STFC, Harwell Campus,
Didcot, OX11 OQX, UK.}}
\newcommand{\oxford}{\affiliation{Department of Physics, University of Oxford, Parks Road, Oxford,
OX1 3PU, UK.}}
\newcommand{\liverpool}{\affiliation{Department of Physics, University of Liverpool, Merseyside, L69 7ZE, UK}}
\newcommand{\dares}{\affiliation{Daresbury Laboratory, UKRI--STFC, Warrington, WA4 4AD, UK.}}

\begin{document}

\title{A High-Flux Source of Cold Strontium with a Loading Rate of $4 \times 10^{10}$ atoms/s for Open Release}

\author{Thomas Walker}
\altaffiliation{These authors contributed equally}
\icl

\author{Anna L.\ Marchant}
\altaffiliation{These authors contributed equally}
\ral

\author{Elliot Bentine}
\oxford
\author{Oliver Buchmueller}
\icl
\author{Katherine Clarke}
\dares
\author{Christopher Foot}
\oxford

\author{Leonie Hawkins}
\icl
\author{Kenneth M.\ Hughes}
\oxford
\author{Kamran Hussain}
\ral
\liverpool
\author{Ludovico Iannizzotto-Venezze}
\icl
\author{Alice Josset}
\icl
\author{Hamza Labiad}
\ral
\author{Dillen Lee}
\icl

\author{Timothy C.\ Thornton-Sparkes}
\ral
\author{Tristan Valenzuela}
\ral
\author{Maurits van der Grinten}
\ral
\author{Andrew Vick}
\dares

\author{Mark G.\ Bason}
\ral
\author{Charles F.\ A.\ Baynham}
\icl
\author{Richard Hobson}
\icl

\date{\today}

\newpage
\begin{abstract}

\noindent

We present a high-flux source of cold strontium atoms based on a two-dimensional magneto-optical trap (2D MOT) and a Zeeman slower. We use the source to load a 3D MOT in a separate science chamber, observing a loading rate of \qty{4e10}{atoms\per\second}---to our knowledge, the highest reported loading flux for strontium. To characterise the vacuum pressure in the science chamber, we load the atoms into a magnetic trap and measure a lifetime of between 8 and 24 seconds, depending on oven temperature. Finally, we characterise the atom flux and velocity distributions from the oven and from the 2D MOT source, finding reasonable agreement with models in the free molecular flow regime. Our results show it is possible to readily produce a cold strontium flux at comparable levels to alkali species, at oven temperatures compatible with long-term operation, and at vacuum pressures suitable for state-of-the-art quantum experiments. We make our design available at no cost, to benefit researchers in the quantum community.

\end{abstract}

\maketitle


\section{Introduction}%
\label{sec:introduction}

Cold atoms have found a broad range of applications from precision metrology with optical clocks \cite{Ludlow2015, Beloy2021, Takamoto2005} to tweezer arrays for quantum computing and simulation \cite{Browaeys2020, Kaufman2021, Manetsch2025}. There is also a growing interest in the application of cold atom quantum technology for fundamental physics, such as long-baseline atom interferometers designed to search for dark matter and gravitational waves \cite{Badurina2020, Abe2021, Canuel2018, Zhan2020, Canuel2020} as well as tests of the equivalence principle \cite{Fray2004, Asenbaum2020, Zhou2022}.

With this development of cold-atom quantum sensors and optical clocks, there is an increasing need for compact, high flux sources. As many experiments target operation close to the standard quantum limit \cite{Baynham2025}, high atom numbers are crucial to enhancing the signal to noise ratio achievable. In the case of alkaline-earth elements, such as strontium, the additional complexity of low vapour pressures at room temperature means that atoms must be heated before any laser cooling stages can be performed. Typically, Zeeman slowing beams are used to decelerate the high-velocity atoms before they are captured in a magneto-optical trap (MOT). The efficiency of this process can be enhanced using multi-frequency Zeeman beams \cite{Li2022, Feng2024}, adding sidebands to the MOT cooling light \cite{Barbiero2020} or deflecting the atomic beam \cite{Yang2015, Li2023}. Compact systems have also been realised by using the residual field from the 2D MOT as the Zeeman slowing field \cite{Nosske2017}. 

Most experiments require additional cooling stages, with atoms from the 2D MOT typically being loaded into a 3D trap. Here, we demonstrate a cold atom flux sufficient to load $4\times 10^{10}$~atoms/s from a source chamber into a 3D MOT, whilst maintaining a good vacuum lifetime, sufficient to reach quantum degeneracy with further cooling \cite{Stellmer2009}. This result is the highest flux reported for a strontium system; our measured value is slightly higher than that in Ref.~\cite{Courtillot2003} but obtained with the oven operating at a considerably reduced temperature, and hence longer operational lifetime for a given load of strontium metal. It also demonstrates cold strontium flux levels on a par with more commonly used alkali systems such as rubidium \cite{Ravenhall2021, Schoser2002, Park2012, Jollenbeck2011}, typically chosen for their relatively lower experimental complexity and high atom numbers. 

Our solution has been designed to scale to production of multiple systems with the needs of long-baseline atom interferometers in mind~\cite{centralized-production}. High-quality engineering drawings mean that production can be mostly out-sourced, reducing the burden of commissioning new experiments. We encourage groups considering a new cold-atom source to contact us --- we will be happy to share our CAD designs to support researchers in the quantum technologies community. 

\section{Setup}
\label{sec:setup}
\subsection{Chamber and 2D MOT}

\begin{figure*}
    \centering
    \includegraphics[width=0.8\linewidth]{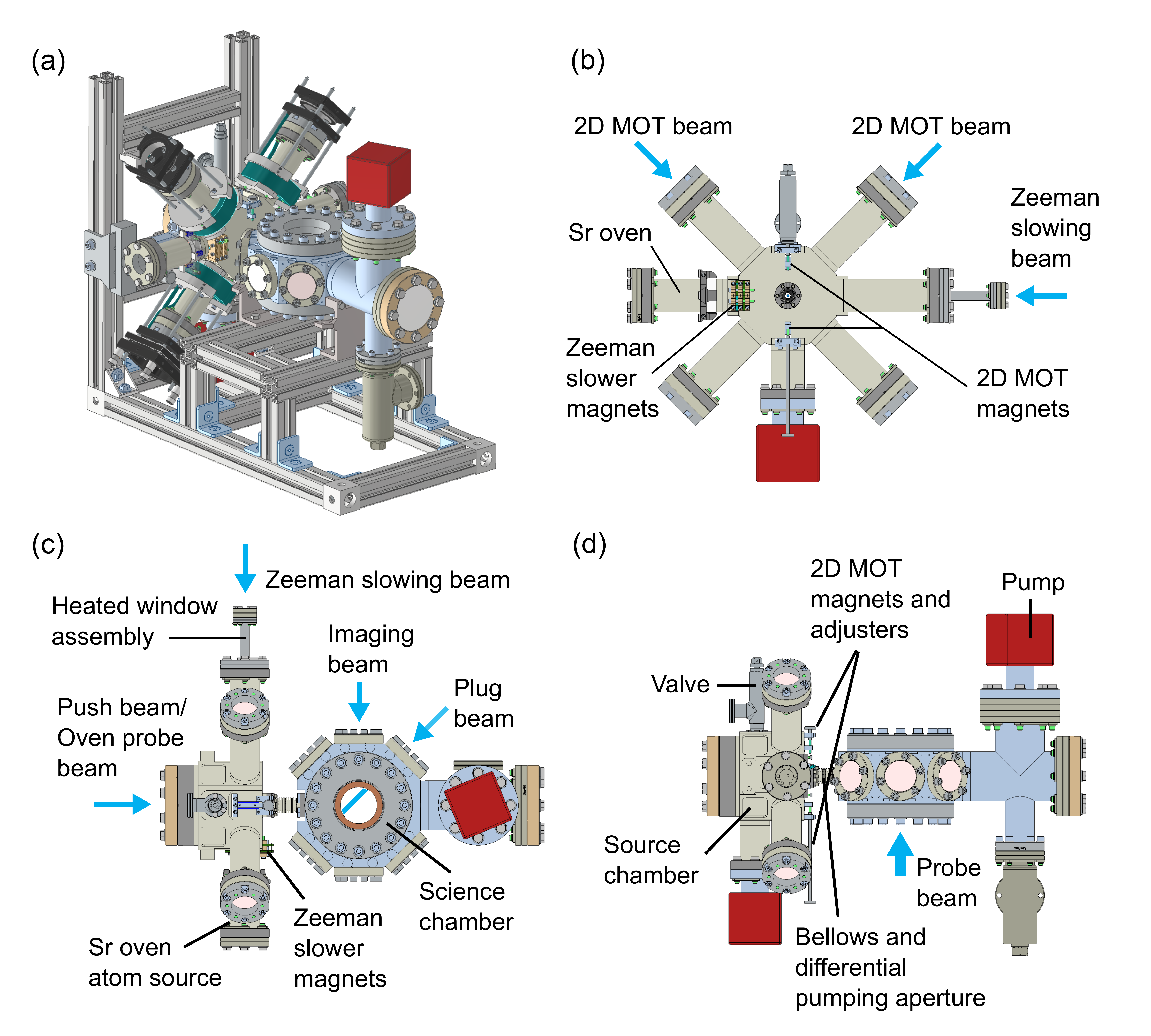}
    \caption{(a) Vacuum system and mounting frame. Some of the 2D MOT optics are shown mounted to the source chamber limbs. The full enclosure forms a full cuboid (670~mm (W) $\times$ 600~mm (D) $\times$ 545~mm (H)) enclosing the chamber, but the top and front faces are not shown here. (b) Source chamber as viewed from the science chamber, showing the location of the magnet assemblies and orientation of the 2D MOT and Zeeman slowing beams. (c) Top view of the vacuum system showing the orientation of the push, Zeeman slowing, oven fluorescence and plug beams used for flux characterisation (see text). (d) Side view of the vacuum system showing the orientation of the probe beam used for flux characterisation in the science chamber.}
    \label{fig:ChamberCAD}
\end{figure*}

We characterise two copies of the vacuum system, one at Imperial College London and one at the Rutherford Appleton Laboratory. The overall system design consists of two chambers: a source chamber, where the 2D MOT is formed, which houses the strontium oven and heated viewport assembly, and a science chamber, used here for diagnostics and to form the 3D MOT. The chambers and mounting frame are shown in Fig.~\ref{fig:ChamberCAD}(a). The two chambers are connected via a flexible bellows containing a differential pumping aperture, 58~mm long with a 7~mm inner diameter, blackened with water-based graphene paint (EM-Tec C32) to minimise stray reflections.

The strontium oven is a modified version of the design presented in \cite{Oxfordoven}, with a larger crucible and a steel nozzle. The oven is heated using six cartridge heaters, inserted into recesses around the oven body. Strontium atoms effuse from the oven through the stainless steel nozzle drilled with 475 circular channels of \nozzleDiameter{} diameter, \nozzleLength{} long, to form a collimated atomic beam. Opposite the oven is a heated sapphire window used for optical access for a Zeeman slowing beam, see Fig.~\ref{fig:ChamberCAD}(b). This window is mounted fully in-vacuum \cite{Oxfordoven} and is heated to avoid it becoming coated with strontium metal. A typical oven operating temperature of \qty{460}{\degreeCelsius} is achieved using a current of \qty{\sim0.4}{\ampere} per heater cartridge and total power of \qty{\sim30}{\watt}. The window is typically heated to \qty{510}{\degreeCelsius} using \qty{\sim53}{\watt} of heater power, but temperatures between 350 and \qty{510}{\degreeCelsius} have been observed to keep the window sufficiently transmissive after months of operation. Permanent magnets are used to generate both the Zeeman slower and 2D MOT fields. Adjustable magnet assemblies used to generate the 2D MOT field gradient are recessed into the top and bottom of the source chamber. The residual field from these magnets contributes to the Zeeman slowing field (see section~\ref{sec:ZS}), but additional, adjustable magnets are also mounted around the oven body to produce the full Zeeman slower field profile. Four coils, wound from aluminium tape, are attached to the source chamber limbs to null any stray fields in the 2D MOT region.    

The 2D MOT is formed by two orthogonal, retro-reflected, circularly polarised beams, red-detuned from the $^1\textrm{S}_0$ to $^1\textrm{P}_1$ transition in $^{88}$Sr.  
Atoms emitted from the oven are first slowed by the Zeeman beam, 
before being captured by the 2D MOT. From there, atoms are transferred into the science chamber using a resonant push beam, which propagates along the chamber axis. The parameters used in the experimental setups are shown in Table~\ref{tab:expnumbers}. 

\begin{table*}[ht]
    \centering
    \begin{tabular}{p{7cm} p{3cm} p{3cm}}
        \toprule
        \textbf{Parameter}  & \textbf{Rutherford} & \textbf{Imperial}  \\
        \midrule
        2D MOT power (single beam)    & \qty{84}{\milli\watt}   & \qty{184}{\milli\watt}  \\
        2D MOT radius                 & 17~mm    & \qty{15}{\milli\meter}  \\
        2D MOT detuning               & -40~MHz & \qty{-55}{\mega\hertz}* \\
        Zeeman slower power           & 68~mW   & \qty{87}{\milli\watt}  \\
        Zeeman slower radius          & 3.5~mm  & \qty{3.2}{\milli\meter}  \\
        Zeeman slower detuning        & -260~MHz& \qty{-300}{\mega\hertz}  \\
        Push beam power               & 0.12~mW & \qty{0.45}{\milli\watt}  \\
        Push beam radius              & 1.3~mm  & \qty{2.5}{\milli\meter}   \\
        Push beam detuning            & On resonance & On resonance  \\
        \bottomrule
    \end{tabular}
    
    \caption{Typical experimental parameters for the Rutherford and Imperial setups. All frequency detunings are relative to the $^1\textrm{S}_0$ to $^1\textrm{P}_1$ transition~\cite{Nosske2017} in $^{88}$Sr, $I_\textrm{sat}$ = 43~mW/cm$^{-2}$. *Modulated with sidebands at $\pm$ \qty{20}{\mega\hertz} (see \cref{sec:3D-mot})
    }
    
    \label{tab:expnumbers}
\end{table*}

\subsection{Zeeman Slower}
\label{sec:ZS}

\begin{figure}[t!]
    \centering
    \begin{subfigure}[b]{\columnwidth}
    \centering
    \includegraphics[width=1\columnwidth]{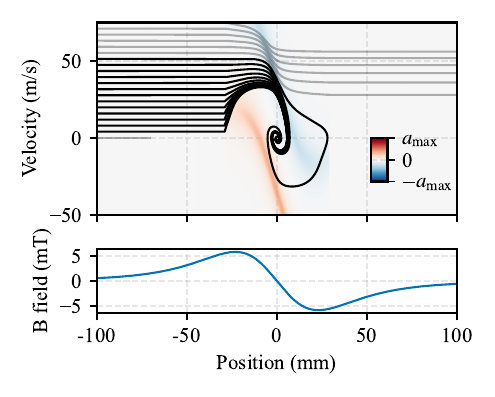}
    \end{subfigure}%
    
    \begin{subfigure}[b]{\columnwidth}
    \centering
    \includegraphics[width=1\columnwidth]{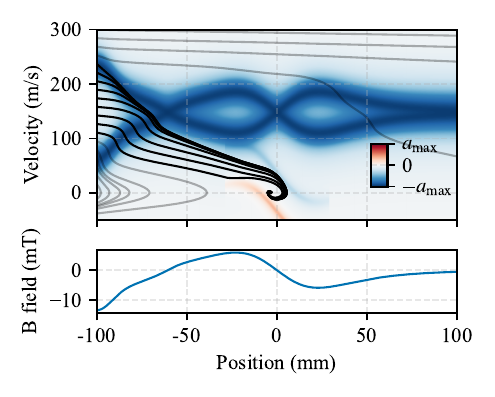}
    \end{subfigure}
    \caption{Simulated trajectories of atoms along the axis of the Zeeman slower.  Upper and lower are without and with Zeeman slower magnets and beam. The heatmap shows the acceleration of the atoms at each position and velocity on a symmetric logarithmic scale, relative to the maximum acceleration $a_{\rm max}$ (see \cref{sec:zeeman-slower-simulations}) from the  $\sigma^+$ and $\sigma^-$ polarised slowing light and MOT beams. Each line is a simulated atom trajectory, with 20 trajectories shown on each plot. Black trajectories are those which are captured by the MOT, grey are those which escape. 
    Below each simulation plot is the calculated transverse field profile along the oven axis.}%
    \label{fig:zs-sim}
\end{figure}

The 2D MOT magnets cause a residual field transverse to the oven axis (see \cref{fig:zs-sim} lower left). This can be used for Zeeman slowing with a beam linearly polarised orthogonal to the  field, such that $\sigma^+$ and $\sigma^-$ transitions are driven. This field is supplemented by four rings of four permanent magnets, each arranged in square around the oven pipe in a Halbach-like configuration (\cref{fig:zs-sim} lower right). Each ring consists of two upward-poled and two downward-poled magnets, with magnets in opposing corners poled in the same direction. This results in a highly uniform vertical field transverse to the atom flux, which falls off quickly outside the Zeeman slowing volume in all directions, to limit the residual field in the 2D MOT and science chamber. The three rings closest to the 2D MOT are \qty{3}{\milli\meter} Y30 ferrite cubes, positioned \qty{15}{\milli\meter} from the oven axis and \numlist[list-final-separator={, and }]{65;72.5;80} \unit{\milli\meter} laterally from the 2D MOT centre. A ring of \dims{10}{5}{2}{\milli\meter} N45H neodymium magnets is positioned \qty{100}{\milli\meter} from the 2D MOT centre, just in front of the oven nozzle, and spaced \qty{23}{\milli\meter} from the oven axis. The magnets are held in place by 3D printed mounts, and can be translated along the oven pipe to fine-tune the field profile, and fixed by set screws.

There is \zsPower{} available for the Zeeman slower beam, which is \zsDetuning{} red-detuned from the $^1\textrm{S}_0 \rightarrow ^1\textrm{P}_1$ transition, with a beam waist ($1/e^2$ radius) of \zsBeamWaist{} at the position of the 2D MOT. 
An offset in the position of the centre of the 2D MOT of \qty{\sim3.8}{\milli\meter} can be seen in Figure \ref{fig:zs-sim} due to the radiation pressure from the Zeeman slower beam. We observe a dimple of similar size in the 2D MOT cloud when the Zeeman slower beam is switched on.

The design of the Zeeman slower was guided by classical trajectory simulations of atoms travelling from the oven to the 2D MOT. The position and velocity of the atoms from the oven are calculated iteratively, considering 1D deceleration along the oven axis due to the MOT and Zeeman slower laser beams, from which a capture efficiency is extracted.  For details see Appendix \ref{sec:zeeman-slower-simulations}. Figure \ref{fig:zs-sim} shows the results of the simulation using experiment parameters measured after the system was built.
The 2D MOT alone has a maximum capture velocity of approximately \qty{55}{\meter\per\second}, covering \qty{0.35}{\percent} of atoms from the oven based on the modified Maxwell-Boltzmann distribution we assume for the oven flux (see Appendix \ref{sec:oven-model}). The Zeeman slower increases this to \qty{255}{\meter\per\second}, at the cost of turning away atoms with velocities below \qty{100}{\meter\per\second}. This range would theoretically allow us to capture up to \qty{15}{\percent} of the atoms in the absence of other losses.

An important consideration  in laser cooling of strontium is the decay from the excited state $^1$P$_1$ to the $^1$D$_2$ state, which takes atoms out of the slowing cycle; with a branching ratio of 1:\num{3.6(3)e4} compared to decay back to the ground state\,\cite{okamotoDirectMeasurement5s5p^1P_12025}. The recoil velocity of strontium atoms is \qty{0.01}{\meter\per\second} (for the \qty{461}{\nano\meter} transition) hence slowing from an initial velocity of \qty{200}{\meter\per\second} requires around \num{2e4} photons, and thus a significant fraction of atoms are lost from the cooling cycle. The simulation predicts that \qty{30}{\percent} of atoms are shelved before reaching the 2D MOT region, and are treated as lost. This estimate, based on Ref.  \cite{okamotoDirectMeasurement5s5p^1P_12025}, shows that making a longer Zeeman slower is unlikely to be useful.

Accounting for losses due to shelving and the transverse velocity distribution of the atoms, the simulations predict a total capture efficiency of \qty{0.15}{\percent} of the oven flux with just the 2D MOT, increasing to \qty{4.1}{\percent} with the Zeeman slower active.

\section{Results}
\label{sec:results}
\subsection{Oven Flux Characterisation}
\label{sec:oven-flux}

The oven flux is characterised using transverse absorption spectroscopy on the Imperial system. A probe beam of power \ovenProbePower{} and diameter \ovenProbeDiameter{} is aligned orthogonally to the atomic beam, along the 2D MOT axis, through the source chamber and into the science chamber. The power of the probe beam is measured on a photodiode behind the exit viewport, and its frequency is recorded using a wavelength meter while the laser is scanned. 

\begin{figure}[ht]
    \centering
    \includegraphics[width=\columnwidth]{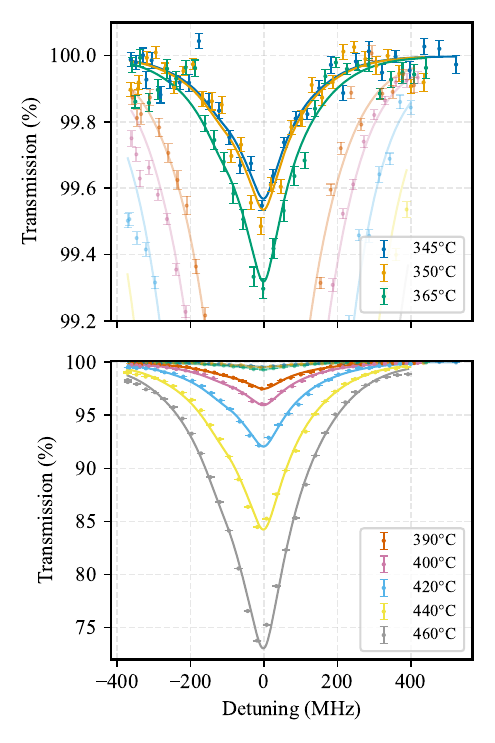}
    \caption{
        Transverse absorption spectrum of the atomic beam from the oven for a range of oven temperatures. The points are experimental data, the solid line is a fit to the model described in the text.
    }%
    \label{fig:oven-spectrum}
\end{figure}

The intensity of transmitted light $I$ is given by the Beer-Lambert law $I=I_0e^{-\mathrm{OD}}$, where the optical density at probe beam detuning $\Delta$ and temperature $T$ is
\begin{equation}
\mathrm{OD}(\Delta, T) = \frac{1}{d}
\int_{-\theta_{\mathrm{lim}}}^{\theta_{\mathrm{lim}}}
\int_{0}^\infty
\, \frac{f(v, T)}{v} \,
\mathcal{I}(\theta)\,
\sigma(\Delta, v, \theta)
\, \, dv \, d\theta.
\label{eq:optical-density}
\end{equation}
$d=\ovenDistance$ is the distance from the oven nozzle to the probe beam, $\theta_{\mathrm{lim}} = \ovenAngleLimitDegree$ is the maximum angle at which an atom can reach the 2D MOT region, $\mathcal{I}(\theta)$ is the angular intensity from the oven, $f(v)$ is the velocity distribution, and $\sigma$ is the absorption cross-section.
See Appendix \ref{sec:oven-model} for details of the model used. We fit the transmission spectra for a range of oven temperatures, shown in Figure~\ref{fig:oven-spectrum}, using this model.
The transmission of the laser varies linearly with laser frequency, even with the oven cold, likely due to parasitic etalon effects from the vacuum viewports. To account for this, we include a linear gradient and a constant offset as free parameters in the fits, which have been subtracted from the data shown in the figure. These parameters are independent for each dataset, to account for the etalons drifting over time.
We find good qualitative agreement between the model and the data, indicating that the model used for oven emission applies across the range of temperatures studied.
There is a slight skew towards lower detunings in the data compared to the model, which may be from a small misalignment of the probe beam, or a nonlinear background dependence on detuning.
With the total oven flux calculated from our oven geometry (see Appendix \ref{sec:oven-model}), \qty{0.58}{\milli\gram} of Sr is consumed per hour at the hottest measured thermocouple temperature of \qty{465}{\degreeCelsius}. At this rate, the \qty{5}{\gram} of Sr originally loaded into our oven would empty after approximately one year of constant operation. 

\subsection{Flux from the 2D MOT}
\label{Sec:2Dflux}

As part of the development of the system at the Rutherford Appleton Laboratory (RAL) we explore the atomic flux generated by the 2D MOT cooling stage. To optimise the cold atom flux from the 2D MOT we measure the arrival of atoms into the science chamber for an oven temperature of 410$^\circ$C. Atoms are detected using fluorescence spectroscopy with a resonant probe, aligned vertically, perpendicular to the atomic beam direction, as shown in Fig.~\ref{fig:ChamberCAD}(d). The beam is retroreflected, giving a total power of 2.1~mW at the position of the atoms, corresponding to an intensity of 0.24~$I_\textrm{sat}$. Fluorescence is collected using a photodiode positioned at $90^\circ$ to both the probe and atomic beams, at the maximum of the dipolar emission pattern, with a $\times 0.5$ telescope between the chamber and photodiode to enhance the collection efficiency.

\begin{figure*}
    \centering
    \includegraphics[width=0.9\linewidth]{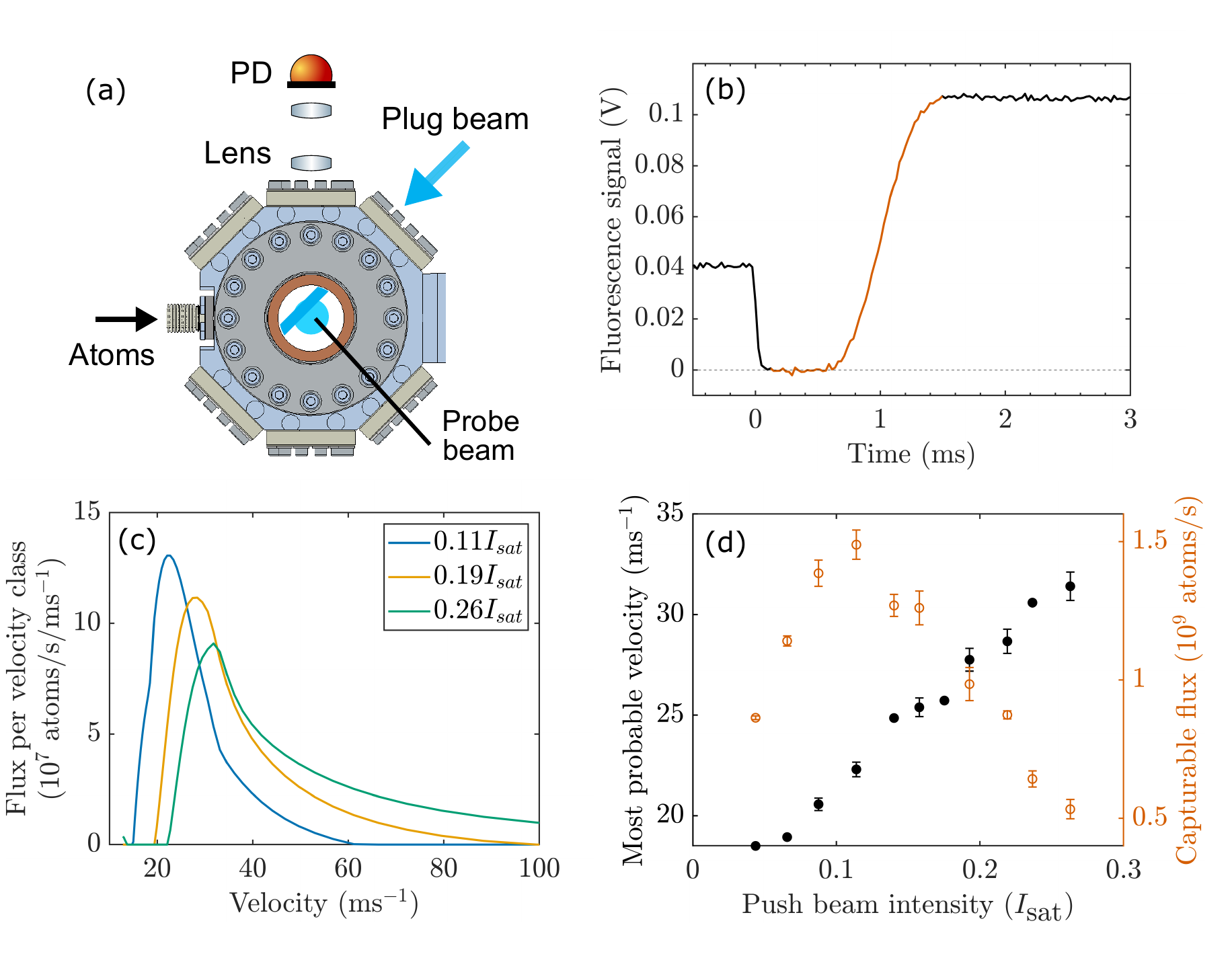}
    \caption{(a) Top view of the science chamber showing the orientation of the plug beam, probe beam (propagating vertically), collection lenses and  photodiode (PD). The black arrow shows the direction in which the atoms propagate. (b) Fluorescence signal used to determine atomic flux. Scatter from the plug beam can be seen up until the light is switched off a $t = 0$~ms. The signal increases as atoms enter the probe region. The region shaded in red is used to generate the curves in panel (c). (c) Example velocity distributions of atoms arriving into the detector region for various push beam intensities. (d) Most probable velocity of atoms reaching the second chamber (filled circles) as a function of push beam intensity, along with the total useful flux (open circles), calculated by integrating the velocity profiles up to 30~ms$^{-1}$.     }
    \label{fig:Flux_characterisation}
\end{figure*}

As the push beam passes along the differential pumping aperture and into the science chamber, atoms are continually scattering photons from the beam and being accelerated. This makes any assumptions about constant velocity of the atoms as they emerge from the 2D MOT region inaccurate. In order to minimise this error we reduce the distance travelled between the end point of the scattering and our detector region. To do this we use a resonant `plug' beam, shown in Fig.~\ref{fig:Flux_characterisation}(a), formed from a Gaussian beam cut in half, aligned with the hard edge close ($\sim2$~\textrm{cm}) to the detection region to prevent atoms initially reaching the probe beam. By extinguishing the plug beam, the atomic signal is effectively switched on and the (almost) fully accelerated atoms are released into the detector region. Making the assumption that the atoms are linearly accelerated (in time) from the source chamber to the science chamber, this gives a somewhat conservative bound of a $<5\%$ error on this method. In practice, the acceleration is likely less than linear as atoms shift out of resonance with the push beam as they accelerate (typical detunings reach $1.5-2 \Gamma$ by the probe region), or simply move out of the push beam path.  

Using a time-of-flight technique \cite{Dieckmann1998}, the gradient of the fluorescence signal switch on can be used to determine the atomic velocity and flux. Using the photodiode response, shown in Fig.~\ref{fig:Flux_characterisation}(b), the flux per velocity class can be calculated as
\begin{equation}
    \Phi(v_x)=\eta\frac{a l}{v_x}\frac{dU(t)}{dt}
    \label{Eq:Fluxpervel}
\end{equation}
where $v_x$ is the longitudinal velocity, $\eta$ is calibration from photodiode voltage to atom number arising from the detection efficiency, $a$ is the effective overlap of the cold atom beam and the probe excitation region, $l$ is the distance travelled by the atoms from the plug beam to the detection region and $U$ is the photodiode voltage. The data are first smoothed with a moving average before the gradient of the photodiode signal is calculated in the region of interest, 0.1~ms $<t<$1.5~ms, as indicated on Fig.~\ref{fig:Flux_characterisation}(b). 

As expected, higher push beam intensities result in a greater acceleration of the travelling atoms and thus increased longitudinal velocities, shown in Fig.~\ref{fig:Flux_characterisation}(c). From these profiles we can extract the most probable velocity of the atoms and a total flux, Fig.~\ref{fig:Flux_characterisation}(d). As the 3D MOT has a finite capture velocity, we integrate Eq.~\ref{Eq:Fluxpervel}, with respect to velocity, up to a limit of 30~ms$^{-1}$ to reflect the total capturable, and therefore useful, flux.

\subsection{3D MOT Loading Rate}
\label{sec:3D-mot}

\begin{figure}
	\includegraphics[width=\columnwidth]{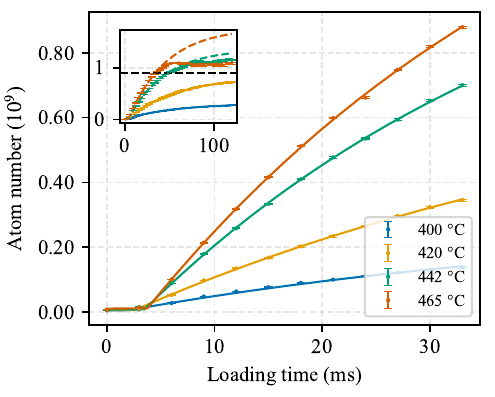}
	\caption{
		3D MOT loading curves for a range of oven temperatures. The points are experimental data, the solid lines are fits to the exponential loading model. The inset shows the full loading curves, with the black dashed line indicating the \qty{0.9e9}{atom} cutoff used for fitting. Uncertainties are standard errors on the mean of 13 measurements.
	}%
	\label{fig:loading-rates}
\end{figure}

To characterise the flux of atoms from the 2D MOT into the 3D MOT, we create a 3D MOT in the science chamber of the Imperial system and measure the loading rate. The 3D MOT field is formed of two pairs of permanent magnet stacks above and below the chamber.
Each stack consists of three \threeDMOTMagnetDims{} N45H Nd magnets. Each 3D MOT beam has a radius of \threeDMOTRadius{} and \threeDMOTPower of power and is operated with a frequency detuning of \threeDMOTDetuning. The 3D MOT is imaged by a CCD camera placed perpendicular to the atom flux from the source chamber, with the imaging beam passing through the opposite viewport for absorption imaging. The loading rate is determined by measuring the number of atoms in the MOT as a function of time after starting to load the MOT. The data is then fitted to a simple model characterised by a loading rate and loss rate (see \cref{sec:3D-mot-methods}). To account for the arrival time of atoms from the 2D MOT to the 3D MOT, we include an onset time and an offset atom number as free parameters in the fit. The results are shown in Figure \ref{fig:loading-rates}.

We observe the MOT becoming unstable at
$\textstyle{\gtrsim}$\qty{e9}{atoms}, with the measured atom number flattening out and becoming noisier (see the inset of Figure \ref{fig:loading-rates}). The MOT also visibly fluctuates in shape close to this atom number. The sharp flattening of the loading curve cannot be accounted for by one- or two-body losses alone.
MOT instability at high atom number has been observed in other systems\,\cite{walkerCollectiveBehaviorOptically1990, pohlSelfdrivenNonlinearDynamics2006, labeyrieSelfSustainedOscillationsLarge2006,gattobigioScalingLawsLarge2010} and is likely caused by a combination of multiphoton scattering and trap nonlinearities. The threshold at which instability occurs could be raised by increasing the detuning of the MOT beams\,\cite{gaudesiusInstabilityThresholdLarge2020}, though we did not explore this in detail.
As we do not have a physical model for this saturation, we only fit data below \qty{0.9e9}{atoms}, where the behaviour is expected to be well described by the simple loading model.

To improve the loading rate into the 3D MOT, repumping light at 707 nm and 679 nm are used to deshelve atoms in the 2D MOT from the $^3$P$_2$ state. We observe an \qty{8}{\percent} increase in loading rate with the repumping light on. We also observe a reduction in the loss rate from the 3D MOT, indicating that some repumping light reaches the 3D MOT.

The 2D MOT beam frequency is modulated at \qty{20}{\mega\hertz} with a modulation index of \qty{0.5}{\radian} to increase the range of addressable velocities\,\cite{andersonEnhancedLoadingMagnetooptic1994,leeOptimizedAtomicFlux2017,barbieroSidebandEnhancedColdAtomic2020}. This leads to an increase in loading rate of \qty{30}{\percent}. Without this effective broadening, we observe the onset of saturation in the loading rate at our maximum available 2D MOT beam power, whereas with the modulation we observe no such saturation.

\begin{figure}
	\includegraphics[width=\columnwidth]{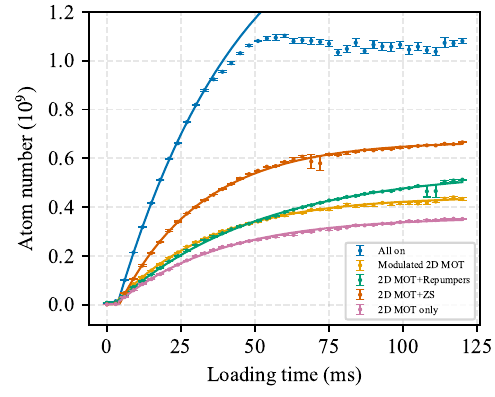}
	\caption{
		Comparison between loading rates for different 2D MOT configurations. The points are experimental data, the solid lines are fits to the exponential loading model. All data shown were taken at \qty{465}{\degreeCelsius} oven temperature.
	}%
	\label{fig:loading-rates-configs}
\end{figure}

A summary of these effects may be seen in Table \ref{tab:loading_rates}, and the loading curves are shown in Figure \ref{fig:loading-rates-configs}. The largest increase in loading rate is due to the Zeeman slower, which increases the loading rate by a factor of \num{2.3}.
Together, there is a factor of \num{4} increase in loading rate with all enhancements active. This is greater than the product of each individual improvement factor, \num{3.2}, meaning there are some synergistic effects when all enhancements are used together. One possible explanation is that the Zeeman slower increases the number of shelved atoms, which would increase the relative impact of the repumping light.
The increase in loading rate from the Zeeman slower is significantly lower than the predicted increase in 2D MOT capture efficiency from the model (around a factor of \num{28}).
Effects not included in the Zeeman slower model which may account for the discrepancy are: the push beam, which would have an impact on the dynamics of the atoms near the 2D MOT; absorption of the MOT and slowing beams, which would reduce the capture efficiency; absorption of the push beam, which would reduce the transfer efficiency to the 3D MOT; and atom-atom collisions. Notably, the relative impact of the effects of absorption and collisions increases for larger 2D MOTs.

\begin{table*}[ht!]
	\centering

	\begin{tabular}{lccc}
		\hline
		\textbf{Configuration}     & \textbf{Loading Rate (\qty{e9}{\per\second})} & \textbf{Loss Rate (\unit{\per\second})} & \textbf{Improvement Factor} \\
		\hline
		2D MOT                     & \num{10.32(8)}                   & \num{28.7(3)}                 & $-$                         \\
		2D MOT + FM & \num{13.47(7)}                   & \num{30.3(3)}                 & \num{1.31(1)}               \\
		2D MOT + RP         & \num{11.2(1)}                    & \num{20.2(3)}                 & \num{1.08(1)}               \\
		2D MOT + ZS     & \num{23.8(2)}                    & \num{35.5(4)}                 & \num{2.30(3)}               \\
		2D MOT + all               & \num{41(2)}                      & \num{23(3)}                   & \num{4.0(2)}                \\
		\hline
	\end{tabular}
	\caption{Comparison of loading and loss rates under different configurations at \qty{465}{\degreeCelsius}. Error bars are parameter uncertainties from the least-squares fit. FM: Frequency modulation. RP: repumping beams.  ZS: Zeeman slower}
	\label{tab:loading_rates}
\end{table*}

\subsection{Magnetic Trap Lifetime}
\label{sec:magnetic}

\begin{figure}
	\includegraphics[width=\columnwidth]{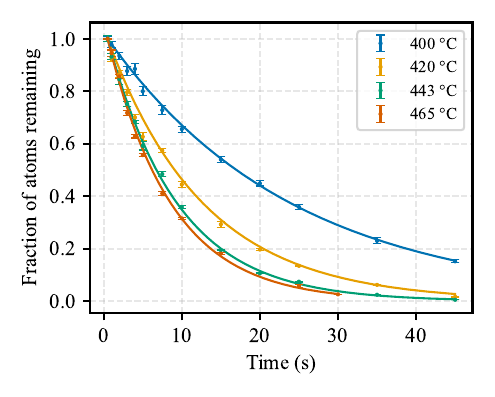}
	\caption{
		Magnetic trap lifetime measurements at different oven temperatures. The points are experimental data, the solid lines are fits to an exponential decay.
		Uncertainties are standard errors on the mean.
	}%
	\label{fig:magnetic-trap-lifetime}
\end{figure}

To characterise the background collision rate at different temperatures, we load magnetic traps in the science chamber and measure the lifetime of atoms in the trap.
The sequence to load the magnetic trap is similar to that used to load the 3D MOT. During the 3D MOT, the repumping beams are off, resulting in some atoms being shelved into the $^3$P$_2$ manifold in the weak-field-seeking state $M_F = 2$. After the 3D MOT, all beams are switched off for a hold time, after which the repumping beams and 3D MOT beams are switched on to detect the fluorescence from the remaining atoms. A plot of the remaining atom fraction against hold time is shown in Figure \ref{fig:magnetic-trap-lifetime}, and the fitted trap lifetimes are given in \cref{tab:loading-lifetimes}. At the highest measured temperature of \qty{465}{\degreeCelsius}, we observe a lifetime of \qty{8.2(2)}{\second}, long enough for evaporative cooling to quantum degeneracy \cite{stellmerBoseEinsteinCondensationStrontium2009}. A long trapping time and high loading flux might be obtainable simultaneously using an atomic oven that can be rapidly heated and cooled\,\cite{Oxfordoven}.

\begin{table*}[ht]
	\centering
	\begin{tabular}{@{}ccc@{}}
		\toprule
		Oven Temperature (\unit{\degreeCelsius}) & Loading Rate ($10^9$/s) & Magnetic Trap Lifetime (s) \\
		\midrule
		400 & \num{6.1(1)}  & \num{23.8(9)} \\
		420 & \num{14.7(1)} & \num{12.4(4)} \\
		440 & \num{33(1)}   & \num{9.0(2)}  \\
		465 & \num{41(2)}   & \num{8.2(2)}  \\
		\bottomrule
	\end{tabular}
	\caption{Loading rates and magnetic trap lifetimes at different oven thermocouple readings. Error bars are parameter uncertainties from the least-squares fit.}%
	\label{tab:loading-lifetimes}
\end{table*}

\section{Conclusion}
We have demonstrated a high flux source of cold strontium, achieved whilst maintaining a vacuum-limited lifetime of more than 8~s, compatible with reaching quantum degeneracy~\cite{Stellmer2009}. We have explored the effectiveness of different techniques to enhance the atom flux from the 2D MOT, such as the introduction of a Zeeman slower, repumping lasers, and frequency modulation of the 2D MOT beams. With all these enhancements, and at the highest thermocouple-measured oven temperature of 465$^\circ$C, we loaded $4\times 10^{10}$~atoms/s into a 3D MOT. We characterised the initial flux from the Sr oven, finding good agreement between our theoretical model and the data, indicating the model's validity over the range of temperatures studied. Comparing with the model, we estimate that the source can operate at highest flux for a year of continuous operation, increasing to 10 years if operated at a 3D MOT loading flux of $6.1\times 10^{9}$~atoms/s. Finally, we studied the velocity distribution of atoms from the 2D MOT source, and we compare different time-of-flight techniques to improve confidence in the measured atom velocities.  Our design is available for researchers in the quantum community to use, free of charge - we encourage interested parties to contact us for details.

In future designs, the slow-atom flux could be improved using more optical power in the 2D MOT and Zeeman slower, or by using additional frequencies and magnetic field shaping for the Zeeman slower, as explored in \cite{Feng2024}. The source and science chamber presented here can be (and has been) transported by crane, interfacing with the $670 \times 600 \times 545$~mm support frame; however, future source designs could further reduce size, weight and power consumption by including in-vacuum mirrors or permanent magnets \cite{bowdenPyramidMOTIntegrated2019, AOSense2025, Li2023}. The oven itself could also be scaled up in size, with a larger volume to hold a larger load of strontium metal (improving oven lifetime) and with an increase in the area of the nozzle (improving flux). The oven flux could be scaled up further using etched, micro-machined, thin fused-silica nozzles, which can be fabricated with $\mathcal{O}(10^5)$ highly-collimating micro-channels---as explored in the same oven design as used in this work \cite{Oxfordoven}.

The high flux, long operational lifetime, and low vacuum pressure of our cold atom source makes it ideal for applications in which flux and reliability are important. 
These results pave the way to high-performance transportable clocks and quantum sensors \cite{Takamoto2020TestGR}, as well as fundamental physics detectors in environments with limited access, such as underground long-baseline atom interferometers \cite{Abe2021,Badurina2020,Baynham2025AICE,Canuel2018}.

\section{Acknowledgements}

We thank Jan Rudolph, Michael Van de Graaff and Jason Hogan for helpful discussions throughout the project. This work was supported by the UKRI International Science Partnerships Fund, and by the USOC grant EP/Y005163/1, awarded by EPSRC, as well as the STFC grant ST/Y004531/1.

\section{Data Availability}
The data that support the findings of this study are available from
the corresponding author upon reasonable request.

\appendix{}
\label{sec:methods}
\section{Oven Flux Model}
\label{sec:oven-model}
In the following, we give formulae for the terms in the equation for optical density in Eqn. \ref{eq:optical-density}. For a full derivation see \cite{paulyAtomMoleculeCluster2000}. 

The angular intensity of the atomic beam, defined as the number of atoms entering the microtubes from the oven per unit time per unit angle, is $\mathcal{I}(\theta) = \mathcal{I}_0 j(\theta)$. The peak angular intensity is related to the total oven flux by $\mathcal{I}_0 = \mathcal{I}(0) = \frac{\Phi}{\pi}$.

The atomic flux at from the oven a temperature $T$ is given by
$\Phi = \frac{1}{4} n \bar{v} A$,
where $\bar{v} = \sqrt{\frac{8 k_B T}{\pi m}}$ is the mean atomic velocity for an atom of mass $m$,  $A$ is the total cross-sectional area of the microtubes, and $n = P/k_B T$, is the atomic number density in the oven,  where the vapour pressure $P$ is given by the empirical formula
$\log_{10}(P)=9.226 + \log_{10}(101325) - 8572 / T - 1.1926 \log_{10}(T)$\,\cite{pucher88SrReferenceData2025}.
Due to collisions with the walls of the microtubes, atoms escape the oven with a probability
$W = 2\int_{0}^{1} j(\theta) \, d(\cos\theta)$.
This gives the expression for the total atomic flux into the chamber
\begin{equation}
	\Phi =  \frac{WPA}{\sqrt{2 \pi m k_B T}}.
\end{equation}

\begin{figure}
	\begin{subfigure}[b]{\columnwidth}
			\centering
			\includegraphics[width=\columnwidth]{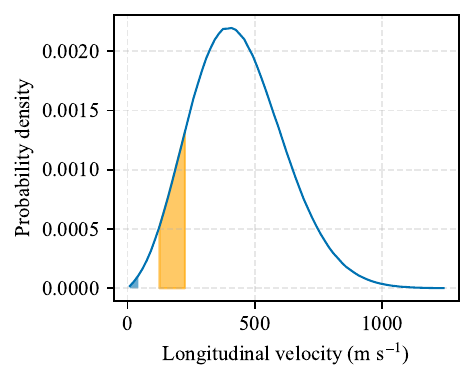}
			\vspace{-2em}
	\end{subfigure}
	\begin{subfigure}[b]{\columnwidth}
			\centering
			\includegraphics[width=\columnwidth, trim={0 1cm 0 1.5cm}, clip]{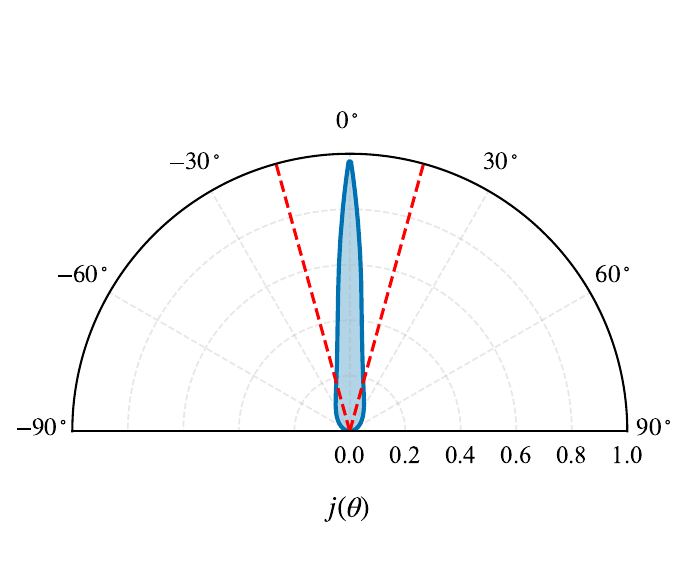}
	\end{subfigure}
	\caption{
		Upper: Longitudinal velocity distribution $f(v)$ for $T=\qty{420}{\degreeCelsius}$, with capture ranges of the 2D MOT (blue shaded) and Zeeman slower (orange shaded) indicated. Lower:
		Angular distribution $j(\theta)$ for our microtube geometry. Red dashed lines indicate the limits on $\theta$ from the chamber geometry.
	}%
	\label{fig:angular-distribution}
\end{figure}

Over the range of temperatures used, our oven microtubes satisfy the condition $\lambda_{\mathrm{MF}} > L > d$, where $\lambda_{\mathrm{MF}} = \frac{k_{B}T}{\pi \sqrt{ 2 }d_{\mathrm{W}}^2P}$ is the mean free path, and $L$ and $d$ are the microtube length and diameter respectively, and $d_{\mathrm{W}}=\qty{498}{\pico\meter}$ is the van der Waals diameter of Sr. From this, we expect the atomic beam to be in the molecular flow regime. In this regime, the angular distribution for a cylinder of length $L$ and radius $r$ is given by
\begin{widetext}
\begin{equation}
	j(\theta) =
	\begin{cases}
		\displaystyle
		\alpha \cos\theta
		+ \frac{2}{\pi} \cos\theta
		\left[
			(1 - \alpha) R(q)
			+ \frac{2}{3q} (1 - 2\alpha)
			\left( 1 - (1 - q^{2})^\frac{3}{2} \right)
			\right],
		 & q \le 1, \\[1.5em]
		\displaystyle
		\alpha \cos\theta
		+ \frac{4}{3\pi q} (1 - 2\alpha) \cos\theta,
		 & q \ge 1,
	\end{cases}
\end{equation}
\end{widetext}
where we have defined the quantities
\begin{align*}
 \beta & = \frac{2r}{L},  \qquad  \qquad  q = \frac{L \, \tan|\theta|}{2r}, \\
	\alpha & = \frac{1}{2}
	- \frac{1}{3 \beta^{2}}
	\frac{
		1 - 2 \beta^{3} + (2\beta^{2} - 1)\sqrt{1+\beta^{2}}
	}{
		\sqrt{1+\beta^{2}} - \beta^{2}\,\sinh^{-1}\left(\frac{1}{\beta}\right)
	}, \\
	R      & = \cos^{-1}(q) - q \sqrt{1-q^{2}}.   
\end{align*}
The two regimes are cases where atoms pass through the microtubes with ($q \le 1$) or without ($q \ge 1$) collisions with the walls. The angular distribution for our microtube geometry is shown in Figure \ref{fig:angular-distribution} (right).

We also assume a modified Maxwell-Boltzmann velocity distribution from the collimated atomic beam
\begin{equation}
	f(v) =  \frac{2v^3}{v_p} e^{-v^2 / v_{p}^2},
\end{equation}
with $v_p = \sqrt{2 k_B T / m}$ being the most probable velocity at temperature $T$. This distribution is shown in Figure \ref{fig:angular-distribution} (left), along with the capture ranges of the 2D MOT and Zeeman slower.

We use a Lorentzian absorption cross-section
\begin{equation}
	\sigma = \frac{\sigma_{0}}{1 + s_0 + \frac{4}{\Gamma^2}(\Delta + kv\sin\theta)^{2}}
\end{equation}
where $s_0=\frac{I}{I_{\mathrm{sat}}}$ is the saturation parameter, $\sigma_0 = \frac{3\lambda^{2}}{2\pi}$ is the resonant cross-section at transition wavelength $\lambda$, $\Delta$ is the detuning of the probe laser from atomic resonance, $k$ is the wavenumber of the probe laser, and $\Gamma$ is the natural linewidth of the transition. For the $^1$S$_0\rightarrow ^1$P$_1$ transition in $^{88}$Sr, $\lambda=\qty{461}{\nano\metre}$, $\Gamma = \qty{32}{\mega\hertz}$, and $I_{\mathrm{sat}} = \qty{43}{\milli\watt\per\centi\meter\squared}$.

\begin{figure}
	\centering
	\includegraphics[width=\columnwidth]{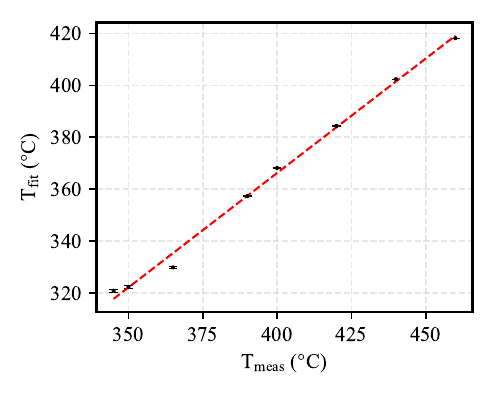}
	\caption{Temperature from fits to the oven flux model $T_{\mathrm{fit}}$ against temperature measured by a thermocouple $T_{\mathrm{meas}}$. The red dashed line is a linear fit with gradient \qty{9(3)}{\degreeCelsius\per\degreeCelsius} and intercept \qty{13(10)}{\degreeCelsius} used to calibrate the measured temperature.}
	\label{fig:temp-calib}
\end{figure}

The fits to the absorption spectra give temperatures which do not exactly match those measured by the thermocouple on the oven body due to temperature gradients across the oven. The temperatures from the model are shown in Figure \ref{fig:temp-calib}. Throughout the paper we have quoted the measured temperature, but the fitted temperature was used for calculation of the Sr lifetime in \cref{sec:oven-flux} and for the Zeeman slower simulation in \cref{sec:zeeman-slower-simulations}.

\section{Zeeman Slower Simulations}
\label{sec:zeeman-slower-simulations}

We consider the 1D motion of atoms along the Zeeman slower axis under the influence of the Zeeman slower and 2D MOT beams and magnetic field. The scattering rate is calculated at each position and velocity, and the resulting acceleration is used to update the atom's velocity and position using a fourth-order Runge-Kutta method.

The scattering rate from a single laser beam is calculated using the standard two-level atom formula
\begin{equation}
	R_{\mathrm{scatt}} = \frac{\Gamma}{2} \frac{s}{1+s},
\end{equation}
where $s =\frac{s_0}{1+(2\Delta_{\mathrm{tot}}/\Gamma)^2}$ is the effective saturation parameter.
The force on the atom from a laser with wavevector $\vec{k}$ is given by $\vec{F} = \hbar \vec{k} R_{\mathrm{scatt}}$.
The total detuning $\Delta_{\mathrm{tot}}$ includes contributions from the laser detuning, Doppler shift, and Zeeman shift due to the magnetic field $\Delta_B = \frac{\Delta m_{F}\mu_{B}B}{\hbar}$ at the atom's position. For $\sigma^{\pm}$ transitions in the $^1$S$_0 \rightarrow ^1$P$_1$ transition, $g_F=1$ and $\Delta m_F = \pm 1$.

Combining multiple beams, the total force on the atom (neglecting multi-photon effects) is given by
\begin{equation}
	\vec{F} = \sum_i \hbar \vec{k}_{i} \frac{\Gamma}{2} \frac{s_i}{1+s_{\mathrm{tot}}}
\end{equation}
with $s_{\mathrm{tot}}=\sum_{j}s_j$. The maximum achievable acceleration is $a_{\mathrm{max}} = \hbar\Gamma/(2m)$.

Only acceleration along the Zeeman slower axis is considered, with the transverse velocity assumed constant. Thus, the vertical confinement of the 2D MOT is neglected. The 3D intensity profiles of the beams are considered, and assumed to be Gaussian. Initial velocities and angles are sampled from the distributions described in Methods \ref{sec:oven-model}.
An atom is considered captured by the 2D MOT if it is within one MOT beam radius of the of the trap centre with a velocity below \qty{10}{\meter\per\second}.
A simple stochastic trajectory method is used to determine the shelving rate, with the dark state scattering probability within a simulation timestep $\delta t$ given by $P_{\mathrm{dark}} = R_{\mathrm{scatt} } \,\mathrm{BR}\,\delta t$, where BR is the branching ratio to the dark state. The capture efficiencies are calculated by simulating \num{10000} trajectories for each configuration.

Magnetic field modelling was performed using the MagPyLib package for Python \cite{ortnerMagpylibFreePython2020} using analytic solutions for fields from permanent cuboid magnets.

\section[]{2D MOT Flux Characterisation Comparison}

\begin{figure}
    \centering
    \includegraphics[width=1\linewidth]{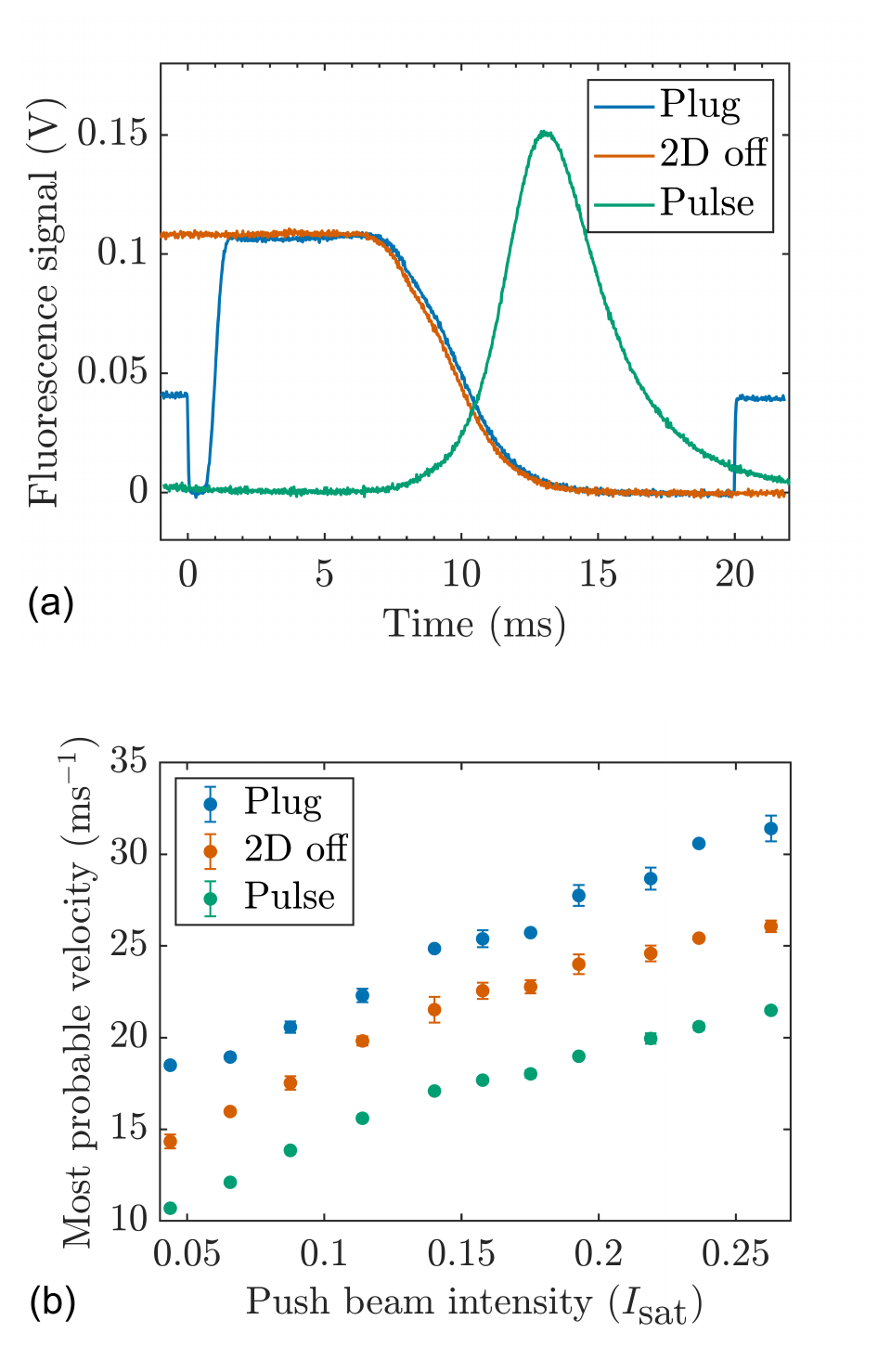}
    \caption{(a) Example fluorescence signals detected in three measurement configurations (see text) for a push beam intensity of $0.07 I_\textrm{sat}$. 0~ms corresponds to the start of the push beam pulse or the switch off of the plug/ 2D MOT light.  (b) Most probable velocity of atoms reaching the probe region, as a function of push beam intensity, for the three methods.  }
    \label{fig:Methods_Vel_comparison}
\end{figure}

To characterise the cold atom flux arriving into the 3D MOT chamber from the 2D MOT we compare three approaches. In all cases atoms are detected with the same resonant probe setup described in section~\ref{Sec:2Dflux}. Example fluorescence traces are shown in Fig.~\ref{fig:Methods_Vel_comparison}(a). 

In the first case, the push beam is pulsed on for 4~ms and the fluorescence of the arriving atoms detected as a function of time. The pulse duration of 4~ms is chosen as beyond this, no change in the signal is observed. Knowing the distance travelled from the 2D MOT source to the detection region, 16~cm, and the arrival time of the atoms it is possible to calculate a velocity profile, using the simplification of an assumed constant acceleration. From this, a most probable velocity can be extracted as a function of push beam intensity. 

The velocity of the atoms can also be determined with another time-of-flight method, instead looking at the decay of the fluorescence signal, as previously demonstrated in \cite{Dieckmann1998}. In this second approach (denoted `2D off' in Fig.~\ref{fig:Methods_Vel_comparison}), the push beam remains constant throughout the sequence but the 2D MOT beams are switched off. This effectively extinguished the atomic source. This method has the advantage of more faithfully reproducing the typical 3D MOT operating conditions of loading via a continuous push beam, accounting for any additional scattering effects. By taking the gradient of the fluorescence decay curve, it is possible to determine the velocity profile of the atoms and the overall flux, as discussed in section~\ref{Sec:2Dflux}.

The continuous scattering of push beam photons by the travelling atoms means the assumption of constant acceleration used in both previous cases gives an underestimate of the arrival velocity. To minimise this effect we also employ a third technique using a resonant `plug' beam, as described previously in section~\ref{Sec:2Dflux}.

A comparison of the most probable velocities obtained from the three techniques is shown in Fig.~\ref{fig:Methods_Vel_comparison}(b). As expected, a higher push beam intensity leads to an increase in most probable velocity as more photons are scattered by the atoms. All methods show this trend, however with varying degrees of inaccuracy. The plug beam method produces the highest velocity values as only maximally accelerated atoms are involved in the rapid switch on signal. We believe this to be our best estimate of the atomic velocity at the position of the 3D MOT, however, due to practical limitations of the setup, it is still an underestimate of the true velocity.

\section{3D MOT Loading}
\label{sec:3D-mot-methods}

The results of the 3D MOT loading rate measurements (see Figure \ref{fig:loading-rates}) are fitted with a simple exponential loading model, where the number of atoms in the MOT $N$ is determined by the loading rate $R$ and loss rate $\gamma$ as

\begin{equation}
	N =
	\begin{cases}
		\displaystyle
		N_0
		 & t < t_0, \\[1.5em]
		\displaystyle
		N_0 + \frac{R}{\gamma}\left(1 - e^{-\gamma (t - t_0)}\right)
		 & t \ge t_0.
	\end{cases}
\end{equation}

The atom number offset $N_0$ and onset time $t_0$ aim to approximate the effect of the atom arrival time distribution without needing to incorporate the distribution analytically in the model.

\section{Magnetic Trap Lifetime}
\label{sec:magnetic-trap-lifetime-methods}

To extract a magnetic trap lifetime from the fluorescence data, we fit a simple exponential decay model to the fluorescence counts $A$ from the camera as a function of hold time $t$ in the magnetic trap,
\begin{equation}
	A = A_0 e^{-\gamma t} + A_{\mathrm{bg}},
\end{equation}
where $A_0$ is the initial fluorescence, $\gamma$ is the loss rate, and $A_{\mathrm{bg}}$ is a constant offset to account for background counts in the imaging. After fitting the background is subtracted and the signal is scaled such that $A_0 = 1$ to give the fractional remaining atom number shown in Figure \ref{fig:magnetic-trap-lifetime}.


\bibliography{hifais, RAL_refs}

@article{centralized-production,
    author = {Stray, B. and Ennis, O. and Hedges, S. and Dey, S. and Langlois, M. and Bongs, K. and Lellouch, S. and Holynski, M. and Bostwick, B. and Chen, J. and Eyler, Z. and Gibson, V. and Harte, T. L. and Hsu, C. C. and Karzazi, M. and Mitchell, J. and Mouelle, N. and Schneider, U. and Tang, Y. and Tkalcec, K. and Zhi, Y. and Clarke, K. and Vick, A. and Bridges, K. and Coleman, J. and Elertas, G. and Hawkins, L. and Hindley, S. and Hussain, K. and Metelko, C. and Throssell, H. and Baynham, C. F. A. and Buchmüller, O. and Evans, D. and Hobson, R. and Iannizzotto-Venezze, L. and Josset, A. and Pasatembou, E. and Sauer, B. E. and Tarbutt, M. R. and Badurina, L. and Beniwal, A. and Blas, D. and Carlton, J. and Ellis, J. and McCabe, C. and Bentine, E. and Booth, M. and Bortoletto, D. and Foot, C. and Gómez-Monedero Castellanos, C. M. and Hird, T. and Hughes, K. and James, A. and Lowe, A. and March-Russell, J. and Schelfhout, J. and Shipsey, I. and Weatherill, D. and Wood, D. and Balashov, S. and Bason, M. G. and Boehm, J. and Courthold, M. and van der Grinten, M. and Majewski, P. and Marchant, A. L. and Newbold, D. and Pan, Z. and Tam, Z. and Valenzuela, T. and Wilmut, I.},
    title = {Centralized design and production of the ultra-high vacuum and laser-stabilization systems for the AION ultra-cold strontium laboratories},
    journal = {AVS Quantum Science},
    volume = {6},
    number = {1},
    pages = {014409},
    year = {2024},
    month = {03},
    issn = {2639-0213},
    doi = {10.1116/5.0172731},
    url = {https://doi.org/10.1116/5.0172731},
}

@misc{Oxfordoven,
  title = {A high-flux atomic strontium oven with light-driven flux modulation},
  author = {Kenneth M. Hughes and Jesse S. Schelfhout and Charu Mishra and Timothy Leese and Elliot Bentine and Christopher J. Foot},
  year = 2026,
  journal = {In preparation}

}

@article{Kaufman2021,
  title = {Quantum Science with Optical Tweezer Arrays of Ultracold Atoms and Molecules},
  author = {Kaufman, Adam M. and Ni, Kang-Kuen},
  year = 2021,
  month = dec,
  journal = {Nature Physics},
  volume = {17},
  number = {12},
  pages = {1324--1333},
  publisher = {Nature Publishing Group},
  issn = {1745-2481},
  doi = {10.1038/s41567-021-01357-2}
}

@article{Manetsch2025,
  title = {A Tweezer Array with 6,100 Highly Coherent Atomic Qubits},
  author = {Manetsch, Hannah J. and Nomura, Gyohei and Bataille, Elie and Lv, Xudong and Leung, Kon H. and Endres, Manuel},
  year = 2025,
  month = nov,
  journal = {Nature},
  volume = {647},
  number = {8088},
  pages = {60--67},
  publisher = {Nature Publishing Group},
  issn = {1476-4687},
  doi = {10.1038/s41586-025-09641-4}
}

@article{Zhou2022,
  title = {Toward a High-Precision Mass--Energy Test of the Equivalence Principle with Atom Interferometers},
  author = {Zhou, Lin and Yan, Si-Tong and Ji, Yu-Hang and He, Chuan and Jiang, Jun-Jie and Hou, Zhuo and Xu, Run-Dong and Wang, Qi and Li, Zhi-Xin and Gao, Dong-Feng and Liu, Min and Ni, Wei-Tou and Wang, Jin and Zhan, Ming-Sheng},
  year = 2022,
  month = dec,
  journal = {Frontiers in Physics},
  volume = {10},
  publisher = {Frontiers},
  issn = {2296-424X},
  doi = {10.3389/fphy.2022.1039119}
}

@article{Canuel2018,
  title = {Exploring Gravity with the {{MIGA}} Large Scale Atom Interferometer},
  author = {Canuel, B. and Bertoldi, A. and Amand, L. and {Pozzo di Borgo}, E. and Chantrait, T. and Danquigny, C. and Dovale {\'A}lvarez, M. and Fang, B. and Freise, A. and Geiger, R. and Gillot, J. and Henry, S. and Hinderer, J. and Holleville, D. and Junca, J. and Lef{\`e}vre, G. and Merzougui, M. and Mielec, N. and Monfret, T. and Pelisson, S. and Prevedelli, M. and Reynaud, S. and Riou, I. and Rogister, Y. and Rosat, S. and Cormier, E. and Landragin, A. and Chaibi, W. and Gaffet, S. and Bouyer, P.},
  year = 2018,
  month = sep,
  journal = {Scientific Reports},
  volume = {8},
  number = {1},
  pages = {14064},
  publisher = {Nature Publishing Group},
  issn = {2045-2322},
  doi = {10.1038/s41598-018-32165-z}
}

@article{Canuel2020,
  title = {{{ELGAR}}---a {{European Laboratory}} for {{Gravitation}} and {{Atom-interferometric Research}}},
  author = {Canuel, B and Abend, S and {Amaro-Seoane}, P and Badaracco, F and Beaufils, Q and Bertoldi, A and Bongs, K and Bouyer, P and Braxmaier, C and Chaibi, W and Christensen, N and Fitzek, F and Flouris, G and Gaaloul, N and Gaffet, S and Garrido Alzar, C L and Geiger, R and {Guellati-Khelifa}, S and Hammerer, K and Harms, J and Hinderer, J and Holynski, M and Junca, J and Katsanevas, S and Klempt, C and Kozanitis, C and Krutzik, M and Landragin, A and L{\`a}zaro Roche, I and Leykauf, B and Lien, Y-H and Loriani, S and Merlet, S and Merzougui, M and Nofrarias, M and Papadakos, P and {Pereira dos Santos}, F and Peters, A and Plexousakis, D and Prevedelli, M and Rasel, E M and Rogister, Y and Rosat, S and Roura, A and Sabulsky, D O and Schkolnik, V and Schlippert, D and Schubert, C and Sidorenkov, L and Siem{\ss}, J-N and Sopuerta, C F and Sorrentino, F and Struckmann, C and Tino, G M and Tsagkatakis, G and Vicer{\'e}, A and {von Klitzing}, W and Woerner, L and Zou, X},
  year = 2020,
  month = oct,
  journal = {Classical and Quantum Gravity},
  volume = {37},
  number = {22},
  pages = {225017},
  publisher = {IOP Publishing},
  issn = {0264-9381},
  doi = {10.1088/1361-6382/aba80e}
}

@article{Zhan2020,
  title = {{{ZAIGA}}: {{Zhaoshan}} Long-Baseline Atom Interferometer Gravitation Antenna},
  shorttitle = {{{ZAIGA}}},
  author = {Zhan, Ming-Sheng and Wang, Jin and Ni, Wei-Tou and Gao, Dong-Feng and Wang, Gang and He, Ling-Xiang and Li, Run-Bing and Zhou, Lin and Chen, Xi and Zhong, Jia-Qi and Tang, Biao and Yao, Zhan-Wei and Zhu, Lei and Xiong, Zong-Yuan and Lu, Si-Bin and Yu, Geng-Hua and Cheng, Qun-Feng and Liu, Min and Liang, Yu-Rong and Xu, Peng and He, Xiao-Dong and Ke, Min and Tan, Zheng and Luo, Jun},
  year = 2020,
  month = mar,
  journal = {International Journal of Modern Physics D},
  volume = {29},
  number = {04},
  pages = {1940005},
  publisher = {World Scientific Publishing Co.},
  issn = {0218-2718},
  doi = {10.1142/S0218271819400054}
}

@article{Fray2004,
  title = {Atomic {{Interferometer}} with {{Amplitude Gratings}} of {{Light}} and {{Its Applications}} to {{Atom Based Tests}} of the {{Equivalence Principle}}},
  author = {Fray, Sebastian and Diez, Cristina Alvarez and H{\"a}nsch, Theodor W. and Weitz, Martin},
  year = 2004,
  month = dec,
  journal = {Physical Review Letters},
  volume = {93},
  number = {24},
  pages = {240404},
  publisher = {American Physical Society},
  doi = {10.1103/PhysRevLett.93.240404}
}

@article{Jollenbeck2011,
  title = {Hexapole-Compensated Magneto-Optical Trap on a Mesoscopic Atom Chip},
  author = {J{\"o}llenbeck, S. and Mahnke, J. and Randoll, R. and Ertmer, W. and Arlt, J. and Klempt, C.},
  year = 2011,
  month = apr,
  journal = {Physical Review A},
  volume = {83},
  number = {4},
  pages = {043406},
  publisher = {American Physical Society},
  doi = {10.1103/PhysRevA.83.043406}
}

@article{Park2012,
  title = {Cold Atomic Beam from a Two-Dimensional Magneto-Optical Trap with Two-Color Pushing Laser Beams},
  author = {Park, Sung Jong and Noh, Jiho and Mun, Jongchul},
  year = 2012,
  month = sep,
  journal = {Optics Communications},
  volume = {285},
  number = {19},
  pages = {3950--3954},
  issn = {0030-4018},
  doi = {10.1016/j.optcom.2012.05.041}
}

@article{Schoser2002,
  title = {Intense Source of Cold {{Rb}} Atoms from a Pure Two-Dimensional Magneto-Optical Trap},
  author = {Schoser, J. and Bat{\"a}r, A. and L{\"o}w, R. and Schweikhard, V. and Grabowski, A. and Ovchinnikov, {\relax Yu}. B. and Pfau, T.},
  year = 2002,
  month = aug,
  journal = {Physical Review A},
  volume = {66},
  number = {2},
  pages = {023410},
  publisher = {American Physical Society},
  doi = {10.1103/PhysRevA.66.023410}
}

@article{Ravenhall2021,
  title = {High-Flux, Adjustable, Compact Cold-Atom Source},
  author = {Ravenhall, Sean and Yuen, Benjamin and Foot, Chris},
  year = 2021,
  month = jul,
  journal = {Optics Express},
  volume = {29},
  number = {14},
  pages = {21143--21159},
  publisher = {Optica Publishing Group},
  issn = {1094-4087},
  doi = {10.1364/OE.423662}
}

@misc{Baynham2025,
  title = {A {{Prototype Atom Interferometer}} to {{Detect Dark Matter}} and {{Gravitational Waves}}},
  author = {Baynham, C. F. A. and Hobson, R. and Buchmueller, O. and Evans, D. and Hawkins, L. and {Iannizzotto-Venezze}, L. and Josset, A. and Lee, D. and Pasatembou, E. and Sauer, B. E. and Tarbutt, M. R. and Walker, T. and Ennis, O. and Chauhan, U. and Brzakalik, A. and Dey, S. and Hedges, S. and Stray, B. and Langlois, M. and Bongs, K. and Hird, T. and Lellouch, S. and Holynski, M. and Bostwick, B. and Chen, J. and Eyler, Z. and Gibson, V. and Harte, T. L. and Hsu, C. C. and Karzazi, M. and Lu, C. and Millward, B. and Mitchell, J. and Mouelle, N. and Panchumarthi, B. and Scheper, J. and Schneider, U. and Su, X. and Tang, Y. and Tkalcec, K. and Zeuner, M. and Zhang, S. and Zhi, Y. and Badurina, L. and Beniwal, A. and Blas, D. and Carlton, J. and Ellis, J. and McCabe, C. and Parish, G. and Govardhan, D. Pathak and Vaskonen, V. and Bowcock, T. and Bridges, K. and Carroll, A. and Coleman, J. and Elertas, G. and Hindley, S. and Metelko, C. and Throssell, H. and Tinsley, J. N. and Bentine, E. and Booth, M. and Bortoletto, D. and Callaghan, N. and Foot, C. and {Gomez-Monedero}, C. and Hughes, K. and James, A. and Leese, T. and Lowe, A. and {March-Russell}, J. and Sander, J. and Schelfhout, J. and Shipsey, I. and Weatherill, D. and Wood, D. and Bason, M. G. and Hussain, K. and Labiad, H. and Marchant, A. L. and Thornton, T. C. and Valenzuela, T. and Balashov, S. N. and Majewski, P. and Newbold, D. and van der Grinten, M. G. D. and Pan, Z. and Tam, Z. and Wilmut, I. and Clarke, K. and Vick, A.},
  year = 2025,
  month = apr,
  number = {arXiv:2504.09158},
  eprint = {2504.09158},
  primaryclass = {hep-ex},
  publisher = {arXiv},
  doi = {10.48550/arXiv.2504.09158},
  archiveprefix = {arXiv}
}

@article{Abe2021,
  title = {Matter-Wave {{Atomic Gradiometer Interferometric Sensor}} ({{MAGIS-100}})},
  author = {Abe, Mahiro and Adamson, Philip and Borcean, Marcel and Bortoletto, Daniela and Bridges, Kieran and Carman, Samuel P and Chattopadhyay, Swapan and Coleman, Jonathon and Curfman, Noah M and DeRose, Kenneth and Deshpande, Tejas and Dimopoulos, Savas and Foot, Christopher J and Frisch, Josef C and Garber, Benjamin E and Geer, Steve and Gibson, Valerie and Glick, Jonah and Graham, Peter W and Hahn, Steve R and Harnik, Roni and Hawkins, Leonie and Hindley, Sam and Hogan, Jason M and Jiang, Yijun and Kasevich, Mark A and Kellett, Ronald J and Kiburg, Mandy and Kovachy, Tim and Lykken, Joseph D and {March-Russell}, John and Mitchell, Jeremiah and Murphy, Martin and Nantel, Megan and Nobrega, Lucy E and Plunkett, Robert K and Rajendran, Surjeet and Rudolph, Jan and Sachdeva, Natasha and Safdari, Murtaza and Santucci, James K and Schwartzman, Ariel G and Shipsey, Ian and Swan, Hunter and Valerio, Linda R and Vasonis, Arvydas and Wang, Yiping and Wilkason, Thomas},
  year = 2021,
  month = jul,
  journal = {Quantum Science and Technology},
  volume = {6},
  number = {4},
  pages = {044003},
  publisher = {IOP Publishing},
  issn = {2058-9565},
  doi = {10.1088/2058-9565/abf719}
}

@article{Asenbaum2020,
  title = {Atom-Interferometric Test of the Equivalence Principle at the ${10}^{\ensuremath{-}12}$ Level},
  author = {Asenbaum, Peter and Overstreet, Chris and Kim, Minjeong and Curti, Joseph and Kasevich, Mark A.},
  year = 2020,
  month = nov,
  journal = {Physical Review Letters},
  volume = {125},
  number = {19},
  pages = {191101},
  publisher = {American Physical Society},
  doi = {10.1103/PhysRevLett.125.191101}
}

@article{Badurina2020,
  title = {{{AION}}: An Atom Interferometer Observatory and Network},
  shorttitle = {{{AION}}},
  author = {Badurina, L. and Bentine, E. and Blas, D. and Bongs, K. and Bortoletto, D. and Bowcock, T. and Bridges, K. and Bowden, W. and Buchmueller, O. and Burrage, C. and Coleman, J. and Elertas, G. and Ellis, J. and Foot, C. and Gibson, V. and Haehnelt, M.G. and Harte, T. and Hedges, S. and Hobson, R. and Holynski, M. and Jones, T. and Langlois, M. and Lellouch, S. and Lewicki, M. and Maiolino, R. and Majewski, P. and Malik, S. and {March-Russell}, J. and McCabe, C. and Newbold, D. and Sauer, B. and Schneider, U. and Shipsey, I. and Singh, Y. and Uchida, M.A. and Valenzuela, T. and van der Grinten, M. and Vaskonen, V. and Vossebeld, J. and Weatherill, D. and Wilmut, I.},
  year = 2020,
  month = may,
  journal = {Journal of Cosmology and Astroparticle Physics},
  volume = {2020},
  number = {05},
  pages = {011},
  issn = {1475-7516},
  doi = {10.1088/1475-7516/2020/05/011}
}

@article{Barbiero2020,
  title = {Sideband-{{Enhanced Cold Atomic Source}} for {{Optical Clocks}}},
  author = {Barbiero, Matteo and Tarallo, Marco G. and Calonico, Davide and Levi, Filippo and Lamporesi, Giacomo and Ferrari, Gabriele},
  year = 2020,
  month = jan,
  journal = {Physical Review Applied},
  volume = {13},
  number = {1},
  pages = {014013},
  publisher = {American Physical Society},
  doi = {10.1103/PhysRevApplied.13.014013}
}

@article{Browaeys2020,
  title = {Many-Body Physics with Individually Controlled {{Rydberg}} Atoms},
  author = {Browaeys, Antoine and Lahaye, Thierry},
  year = 2020,
  month = feb,
  journal = {Nature Physics},
  volume = {16},
  number = {2},
  pages = {132--142},
  publisher = {Nature Publishing Group},
  issn = {1745-2481},
  doi = {10.1038/s41567-019-0733-z}
}

@article{Courtillot2003,
  title = {Efficient Cooling and Trapping of Strontium Atoms},
  author = {Courtillot, I. and Quessada, A. and Kovacich, R. P. and Zondy, J.-J. and Landragin, A. and Clairon, A. and Lemonde, P.},
  year = 2003,
  month = mar,
  journal = {Optics Letters},
  volume = {28},
  number = {6},
  pages = {468--470},
  publisher = {Optica Publishing Group},
  issn = {1539-4794},
  doi = {10.1364/OL.28.000468}
}

@article{Dieckmann1998,
  title = {Two-Dimensional Magneto-Optical Trap as a Source of Slow Atoms},
  author = {Dieckmann, K. and Spreeuw, R. J. C. and Weidem{\"u}ller, M. and Walraven, J. T. M.},
  year = 1998,
  month = nov,
  journal = {Physical Review A},
  volume = {58},
  number = {5},
  pages = {3891--3895},
  publisher = {American Physical Society},
  doi = {10.1103/PhysRevA.58.3891}
}

@article{Feng2024,
  title = {High Flux Strontium Atom Source},
  author = {Feng, C-H and Robert, P and Bouyer, P and Canuel, B and Li, J and Das, S and Kwong, C C and Wilkowski, D and Prevedelli, M and Bertoldi, A},
  year = 2024,
  month = mar,
  journal = {Quantum Science and Technology},
  volume = {9},
  number = {2},
  pages = {025017},
  publisher = {IOP Publishing},
  issn = {2058-9565},
  doi = {10.1088/2058-9565/ad310b}
}

@article{Li2022,
  title = {Bi-Color Atomic Beam Slower and Magnetic Field Compensation for Ultracold Gases},
  author = {Li, Jianing and Lim, Kelvin and Das, Swarup and {Zanon-Willette}, Thomas and Feng, Chen-Hao and Robert, Paul and Bertoldi, Andrea and Bouyer, Philippe and Kwong, Chang Chi and Lan, Shau-Yu and Wilkowski, David},
  year = 2022,
  month = dec,
  journal = {AVS Quantum Science},
  volume = {4},
  number = {4},
  pages = {046801},
  issn = {2639-0213},
  doi = {10.1116/5.0126745}
}

@article{Li2023,
  title = {An Integrated High-Flux Cold Atomic Beam Source for Strontium},
  author = {Li, Jie and Jia, Zhi-Peng and Liu, Peng and Liu, Xiao-Yong and Wang, De-Zhong and Kong, De-Quan and Li, Su-Peng and Cui, Xing-Yang and Dai, Han-Ning and Chen, Yu-Ao and Pan, Jian-Wei},
  year = 2023,
  month = sep,
  journal = {Review of Scientific Instruments},
  volume = {94},
  number = {9},
  pages = {093202},
  issn = {0034-6748},
  doi = {10.1063/5.0162128}
}

@article{Ludlow2015,
  title = {Optical Atomic Clocks},
  author = {Ludlow, Andrew D. and Boyd, Martin M. and Ye, Jun and Peik, E. and Schmidt, P. O.},
  year = 2015,
  month = jun,
  journal = {Reviews of Modern Physics},
  volume = {87},
  number = {2},
  pages = {637--701},
  publisher = {American Physical Society},
  doi = {10.1103/RevModPhys.87.637}
}

@article{Nosske2017,
  title = {Two-Dimensional Magneto-Optical Trap as a Source for Cold Strontium Atoms},
  author = {Nosske, Ingo and Couturier, Luc and Hu, Fachao and Tan, Canzhu and Qiao, Chang and Blume, Jan and Jiang, Y. H. and Chen, Peng and Weidem{\"u}ller, Matthias},
  year = 2017,
  month = nov,
  journal = {Physical Review A},
  volume = {96},
  number = {5},
  pages = {053415},
  publisher = {American Physical Society},
  doi = {10.1103/PhysRevA.96.053415}
}

@article{Stellmer2009,
  title = {Bose-{{Einstein Condensation}} of {{Strontium}}},
  author = {Stellmer, Simon and Tey, Meng Khoon and Huang, Bo and Grimm, Rudolf and Schreck, Florian},
  year = 2009,
  month = nov,
  journal = {Physical Review Letters},
  volume = {103},
  number = {20},
  pages = {200401},
  publisher = {American Physical Society},
  doi = {10.1103/PhysRevLett.103.200401}
}

@article{Yang2015,
  title = {A High Flux Source of Cold Strontium Atoms},
  author = {Yang, Tao and Pandey, Kanhaiya and Pramod, Mysore Srinivas and Leroux, Frederic and Kwong, Chang Chi and Hajiyev, Elnur and Chia, Zhong Yi and Fang, Bess and Wilkowski, David},
  year = 2015,
  month = oct,
  journal = {The European Physical Journal D},
  volume = {69},
  number = {10},
  pages = {226},
  issn = {1434-6079},
  doi = {10.1140/epjd/e2015-60288-y}
}

@Article{Takamoto2005,
author={Takamoto, Masao
and Hong, Feng-Lei
and Higashi, Ryoichi
and Katori, Hidetoshi},
title={An optical lattice clock},
journal={Nature},
year={2005},
month={May},
day={01},
volume={435},
number={7040},
pages={321-324},
issn={1476-4687},
doi={10.1038/nature03541}
}

@Article{Beloy2021,
author={Beloy, Kyle
and Bodine, Martha I.
and Bothwell, Tobias
and Brewer, Samuel M.
and Bromley, Sarah L.
and Chen, Jwo-Sy
and Desch{\^e}nes, Jean-Daniel
and Diddams, Scott A.
and Fasano, Robert J.
and Fortier, Tara M.
and Hassan, Youssef S.
and Hume, David B.
and Kedar, Dhruv
and Kennedy, Colin J.
and Khader, Isaac
and Koepke, Amanda
and Leibrandt, David R.
and Leopardi, Holly
and Ludlow, Andrew D.
and McGrew, William F.
and Milner, William R.
and Newbury, Nathan R.
and Nicolodi, Daniele
and Oelker, Eric
and Parker, Thomas E.
and Robinson, John M.
and Romisch, Stefania
and Sch{\"a}ffer, Stefan A.
and Sherman, Jeffrey A.
and Sinclair, Laura C.
and Sonderhouse, Lindsay
and Swann, William C.
and Yao, Jian
and Ye, Jun
and Zhang, Xiaogang
and Collaboration*, Boulder Atomic Clock Optical Network (BACON)},
title={Frequency ratio measurements at 18-digit accuracy using an optical clock network},
journal={Nature},
year={2021},
month={Mar},
day={01},
volume={591},
number={7851},
pages={564-569},
issn={1476-4687},
doi={10.1038/s41586-021-03253-4}
}

@misc{AOSense2025,
    author = {AOSense},
    howpublished = {\url{https://aosense.com/products/atom-beam-sources/cold-atomic-beam-system/}} 
}

@article{Baynham2025AICE,
  title        = {Letter of Intent: {AICE} -- 100m Atom Interferometer Experiment at {CERN}},
  author       = {Baynham, Charles and Bertoldi, Andrea and Blas, Diego and Buchmueller, Oliver and Calatroni, Sergio and Charmandaris, Vassilis and Chiofalo, Maria Luisa and Cladé, Pierre and Coleman, Jonathon and Di Pumpo, Fabio and Ellis, John and Gaaloul, Naceur and Guellati-Khelifa, Saïda and Harte, Tiffany and Hobson, Richard and Holynski, Michael and Lellouch, Samuel and Lombriser, Lucas and Lopez Asamar, Elias and Maggiore, Michele and McCabe, Christopher and Mitchell, Jeremiah and Rasel, Ernst M. and Sanchez Nieto, Federico and Schleich, Wolfgang and Schlippert, Dennis and Schneider, Ulrich and Schramm, Steven and Soares-Santos, Marcelle and Tino, Guglielmo M. and Tinsley, Jonathan N. and Valenzuela, Tristan and van der Grinten, Maurits and von Klitzing, Wolf},
  journal      = {arXiv preprint arXiv:2509.11867},
  year         = {2025},
  doi          = {10.48550/arXiv.2509.11867}
}

@article{Takamoto2020TestGR,
  title        = {Test of general relativity by a pair of transportable optical lattice clocks},
  author       = {Takamoto, Masao and Ushijima, Ichiro and Ohmae, Noriaki and Yahagi, Toshihiro and Kokado, Kensuke and Shinkai, Hisaaki and Katori, Hidetoshi},
  journal      = {Nature Photonics},
  volume       = {14},
  number       = {7},
  pages        = {411--415},
  year         = {2020},
  doi          = {10.1038/s41566-020-0619-8},
  url          = {https://www.nature.com/articles/s41566-020-0619-8}
}

@article{andersonEnhancedLoadingMagnetooptic1994,
  title = {Enhanced Loading of a Magneto-Optic Trap from an Atomic Beam},
  author = {Anderson, Brian P. and Kasevich, Mark A.},
  year = 1994,
  month = nov,
  journal = {Physical Review A},
  volume = {50},
  number = {5},
  pages = {R3581-R3584},
  issn = {1050-2947, 1094-1622},
  doi = {10.1103/PhysRevA.50.R3581},
  urldate = {2025-11-07},
  copyright = {http://link.aps.org/licenses/aps-default-license},
  langid = {english}
}

@article{barbieroSidebandEnhancedColdAtomic2020,
  title = {Sideband-{{Enhanced Cold Atomic Source}} for {{Optical Clocks}}},
  author = {Barbiero, Matteo and Tarallo, Marco G. and Calonico, Davide and Levi, Filippo and Lamporesi, Giacomo and Ferrari, Gabriele},
  year = 2020,
  month = jan,
  journal = {Physical Review Applied},
  volume = {13},
  number = {1},
  pages = {014013},
  issn = {2331-7019},
  doi = {10.1103/PhysRevApplied.13.014013},
  urldate = {2025-11-07},
  langid = {english}
}

@article{bowdenPyramidMOTIntegrated2019,
  title = {A Pyramid {{MOT}} with Integrated Optical Cavities as a Cold Atom Platform for an Optical Lattice Clock},
  author = {Bowden, William and Hobson, Richard and Hill, Ian R. and Vianello, Alvise and Schioppo, Marco and Silva, Alissa and Margolis, Helen S. and Baird, Patrick E. G. and Gill, Patrick},
  year = 2019,
  month = aug,
  journal = {Scientific Reports},
  volume = {9},
  number = {1},
  pages = {11704},
  publisher = {Nature Publishing Group},
  issn = {2045-2322},
  doi = {10.1038/s41598-019-48168-3},
  urldate = {2025-09-29},
  copyright = {2019 The Author(s)},
  langid = {english}
}

@article{gattobigioScalingLawsLarge2010,
  title = {Scaling Laws for Large Magneto-Optical Traps},
  author = {Gattobigio, G L and Pohl, T and Labeyrie, G and Kaiser, R},
  year = 2010,
  month = feb,
  journal = {Physica Scripta},
  volume = {81},
  number = {2},
  pages = {025301},
  issn = {0031-8949, 1402-4896},
  doi = {10.1088/0031-8949/81/02/025301},
  urldate = {2025-11-07},
  langid = {english}
}

@article{gaudesiusInstabilityThresholdLarge2020,
  title = {Instability Threshold in a Large Balanced Magneto-Optical Trap},
  author = {Gaudesius, M. and Kaiser, R. and Labeyrie, G. and Zhang, Y.-C. and Pohl, T.},
  year = 2020,
  month = may,
  journal = {Physical Review A},
  volume = {101},
  number = {5},
  pages = {053626},
  issn = {2469-9926, 2469-9934},
  doi = {10.1103/PhysRevA.101.053626},
  urldate = {2025-11-07},
  langid = {english}
}

@article{labeyrieSelfSustainedOscillationsLarge2006,
  title = {Self-{{Sustained Oscillations}} in a {{Large Magneto-Optical Trap}}},
  author = {Labeyrie, G. and Michaud, F. and Kaiser, R.},
  year = 2006,
  month = jan,
  journal = {Physical Review Letters},
  volume = {96},
  number = {2},
  pages = {023003},
  publisher = {American Physical Society},
  doi = {10.1103/PhysRevLett.96.023003},
  urldate = {2025-09-29}
}

@article{leeOptimizedAtomicFlux2017,
  title = {Optimized Atomic Flux from a Frequency-Modulated Two-Dimensional Magneto-Optical Trap for Cold Fermionic Potassium Atoms},
  author = {Lee, Jae Hoon and Mun, Jongchul},
  year = 2017,
  month = jul,
  journal = {JOSA B, Vol. 34, Issue 7, pp. 1415-1420},
  publisher = {Optica Publishing Group},
  doi = {10.1364/JOSAB.34.001415},
  urldate = {2025-10-13},
  copyright = {\copyright{} 2017 Optical Society of America},
  langid = {english}
}

@misc{okamotoDirectMeasurement5s5p^1P_12025,
  author = {Okamoto, Naohiro and Aoki, Takatoshi and Torii, Yoshio},
  year = 2025,
  month = oct,
  number = {arXiv:2510.22184},
  eprint = {2510.22184},
  primaryclass = {physics},
  publisher = {arXiv},
  doi = {10.48550/arXiv.2510.22184},
  urldate = {2025-11-06},
  archiveprefix = {arXiv},
  langid = {english},
  title = {Direct {{Measurement}} of the {${5s5p\,{}^1P_1 \to 5s4d\,{}^1D_2}$} {{Decay Rate}} in {{Strontium}}}
}

@article{ortnerMagpylibFreePython2020,
  title = {Magpylib: {{A}} Free {{Python}} Package for Magnetic Field Computation},
  shorttitle = {Magpylib},
  author = {Ortner, Michael and Coliado Bandeira, Lucas Gabriel},
  year = 2020,
  month = jan,
  journal = {SoftwareX},
  volume = {11},
  pages = {100466},
  issn = {2352-7110},
  doi = {10.1016/j.softx.2020.100466},
  urldate = {2025-11-10}
}

@book{paulyAtomMoleculeCluster2000,
  title = {Atom, {{Molecule}}, and {{Cluster Beams I}}},
  author = {Pauly, Hans},
  year = 2000,
  series = {Springer {{Series}} on {{Atomic}}, {{Optical}}, and {{Plasma Physics}}},
  volume = {28},
  publisher = {Springer},
  address = {Berlin, Heidelberg},
  doi = {10.1007/978-3-662-04213-7},
  urldate = {2025-11-07},
  copyright = {http://www.springer.com/tdm},
  isbn = {978-3-642-08623-6 978-3-662-04213-7}
}

@article{pohlSelfdrivenNonlinearDynamics2006,
  title = {Self-Driven Nonlinear Dynamics in Magneto-Optical Traps},
  author = {Pohl, T. and Labeyrie, G. and Kaiser, R.},
  year = 2006,
  month = aug,
  journal = {Physical Review A},
  volume = {74},
  number = {2},
  pages = {023409},
  issn = {1050-2947, 1094-1622},
  doi = {10.1103/PhysRevA.74.023409},
  urldate = {2025-11-07},
  copyright = {http://link.aps.org/licenses/aps-default-license},
  langid = {english}
}

@misc{pucher88SrReferenceData2025,
  author = {Pucher, Sebastian and Kristensen, Sofus Laguna and Kroeze, Ronen M.},
  year = 2025,
  month = aug,
  number = {arXiv:2507.10487},
  eprint = {2507.10487},
  primaryclass = {physics},
  publisher = {arXiv},
  doi = {10.48550/arXiv.2507.10487},
  urldate = {2025-11-10},
  archiveprefix = {arXiv},
  title = {{$^{88}$}Sr {{Reference Data}}}
}

@article{stellmerBoseEinsteinCondensationStrontium2009,
  title = {Bose-{{Einstein Condensation}} of {{Strontium}}},
  author = {Stellmer, Simon and Tey, Meng Khoon and Huang, Bo and Grimm, Rudolf and Schreck, Florian},
  year = 2009,
  month = nov,
  journal = {Physical Review Letters},
  volume = {103},
  number = {20},
  pages = {200401},
  issn = {0031-9007, 1079-7114},
  doi = {10.1103/PhysRevLett.103.200401},
  urldate = {2025-10-31},
  copyright = {http://link.aps.org/licenses/aps-default-license},
  langid = {english}
}

@article{walkerCollectiveBehaviorOptically1990,
  title = {Collective Behavior of Optically Trapped Neutral Atoms},
  author = {Walker, Thad and Sesko, David and Wieman, Carl},
  year = 1990,
  month = jan,
  journal = {Physical Review Letters},
  volume = {64},
  number = {4},
  pages = {408--411},
  issn = {0031-9007},
  doi = {10.1103/PhysRevLett.64.408},
  urldate = {2025-11-07},
  copyright = {http://link.aps.org/licenses/aps-default-license},
  langid = {english}
}

\end{document}